\documentstyle[prl,aps,epsfig,twocolumn,floats]{revtex}

\begin{document}


\wideabs{
\title{The strongly enhanced magnetic excitations near the quantum 
critical point of Cr$_{\text{1-x}}$V$_{\text{x}}$ and why strong 
exchange enhancement need not imply heavy fermion behavior}

\author{S. M. Hayden$^{1}$, R. Doubble$^{1}$, G. Aeppli$^{2}$, 
T. G. Perring$^{3}$, E. Fawcett$^{4}$ }
\address{
$^{1}$H. H. Wills Physics Laboratory, University of Bristol, Tyndall 
Avenue, Bristol BS8 1TL, United Kingdom \\
$^{2}$NEC Research Institute, 4 Independence Way, Princeton, NJ 08540
\\
$^{3}$ISIS Facility, Rutherford Appleton Laboratory, Chilton, Didcot,
OX11 
0QX, United Kingdom \\
$^{4}$Department of Physics, University of Toronto, Toronto, M5S 1A7, 
Canada
}

\date{18 November 1999}
\maketitle

\begin{abstract}
Inelastic neutron scattering reveals strong spin fluctuations with
energies as high as 0.4eV in the nearly antiferromagnetic metal
Cr$_{0.95}$V$_{0.05}$.  The magnetic response is well described by a
modified Millis-Monien-Pines function.  From the low-energy response,
we deduce a large exchange enhancement, more than an order of
magnitude larger than the corresponding enhancement of the
low-temperature electronic heat capacity $\gamma T$.  A scaling
relationship between $\gamma$ and the inverse of the
wavevector-averaged spin relaxation rate $\Gamma_{\text{ave}}$ is
demonstrated for a number of magnetically correlated metals.
\end{abstract}
\pacs{PACS numbers:  75.40.Gb, 75.20.-s, 75.20 En, 61.12.-q}
}

\narrowtext

Many metals close to a magnetic instability at low temperatures display
novel 
behavior in their physical properties, for example a linear temperature
dependence of the resistivity.  Notable examples include 
the lamellar CuO systems\cite{HighTc}, non-Fermi liquid
systems\cite{NFL},
heavy fermions\cite{HF} and nearly ferromagnetic
metals\cite{Bernhoeft89}.  
The proximity of an antiferromagnetic phase is also widely believed to
be 
related to the occurrence of some forms of
superconductivity\cite{Mathur98}. 
Clearly, the magnetic excitations are largely responsible for the
properties of such materials. 
In this paper, we study the magnetic excitations near
the quantum critical point (QCP) of
the structurally simple (body centered cubic) $3d$ metal Cr$_{1-x}$V$_{x}$.  
For $x$ $\leq$ 0.35, Cr$_{1-x}$V$_{x}$ displays antiferromagnetic order 
in the form of an incommensurate spin density wave as in pure
chromium\cite{Cr}. The point ($x$=0.35,$T$=0) in the $x-T$ plane
(see Fig.~1(a)) is a QCP where a phase
transition (the ordering of $3d$ moments) occurs at zero temperature.  

Specifically, we studied the nearly antiferromagnetic composition 
Cr$_{0.95}$V$_{0.05}$. 
Strong magnetic
excitations were observed: in common with $4f$ and $5f$ materials near 
quantum criticality.
However, in
contrast, the excitations in Cr$_{0.95}$V$_{0.05}$ are strong only in a small
portion of reciprocal space. 
Other findings of the present study are: (i)
magnetic excitations exist with energies comparable to those of the
highest frequency spin waves in strong ferromagnets such as
iron\cite{Paul88} and nickel\cite{Mook85}; (ii) a simple
phenomenological response function with four parameters describes the
magnetic response over an extremely wide (0.004-0.4eV) energy range;
(iii) Cr$_{0.95}$V$_{0.05}$ shows a large exchange enhancement of $28
\pm 4$; (iv) there is a universal relationship between the
Brillouin zone (BZ) averaged spin relaxation rate and the electronic
heat capacity for Cr$_{0.95}$V$_{0.05}$ and other magnetically
correlated metals; (v) the low effective mass of the quasiparticles in
Cr$_{0.95}$V$_{0.05}$ which coexists with the high exchange
enhancement is due to the small phase space occupied by the exchange
enhanced fluctuations.
 
Reactor-based neutron scattering measurements have demonstrated\cite{Fawcett88}
incommensurate magnetic correlations in Cr$_{0.95}$V$_{0.05}$ for
thermal energies.  In particular, the low frequency magnetic response
peaks at six symmetry-equivalent positions near the H or (100)-type
point of the BZ.  By using the HET and MARI spectrometers\cite{Hayden97} at 
the ISIS spallation source, we have extended the energy range by almost 
an order of magnitude.
Absolute intensities were obtained via normalization to
measured incoherent scattering from a vanadium standard and coherent
phonon scattering from the sample\cite{Doubble98}. The accuracy
is about $\pm$10\%.  Our sample, from the Materials Preparation Center
of Ames Laboratory, was a 47.9 g arc-melted single crystal.  The
lattice constant at $T$=2 K was $a$=2.895 \AA. Elastic scans performed
using a cold-source three-axis spectrometer ($k_i$=1.5 \AA$^{-1}$)
revealed no antiferromagnetic order for temperatures as low as $T$=2~K.

Neutron scattering measures the imaginary part of the generalized 
magnetic susceptibility $\chi^{\prime\prime}({\bf Q},\omega)$, the 
cross-section for a paramagnet such as Cr$_{0.95}$V$_{0.05}$ 
is,
\begin{eqnarray}
\label{eq1}
\frac{d^2\sigma}{d\Omega \, dE} &=& (\gamma r_{\text{e}})^2
\frac{k_f}{k_i}
\left| F({\bf Q})\right|^2 
\left( 
\frac{2/ \pi g^{2} \mu^{2}_{\rm B}}{1-\exp(-\hbar\omega/kT)}
\right) \nonumber \\
&& \times \chi^{\prime\prime}({\bf Q},\hbar\omega),
\end{eqnarray}
where $(\gamma r_{\text{e}})^2$=0.2905 barn sr$^{-1}$, ${\bf k}_{i}$
and ${\bf k}_{f}$ are the incident and final neutron wavevectors and
$|F({\bf Q})|^2$ is the magnetic form factor.  We label wavevectors by their positions in reciprocal space ${\bf
Q}=(h,k,l)$.  

\begin{figure}[t]
\centering
\epsfxsize=8cm
\mbox{\epsfbox{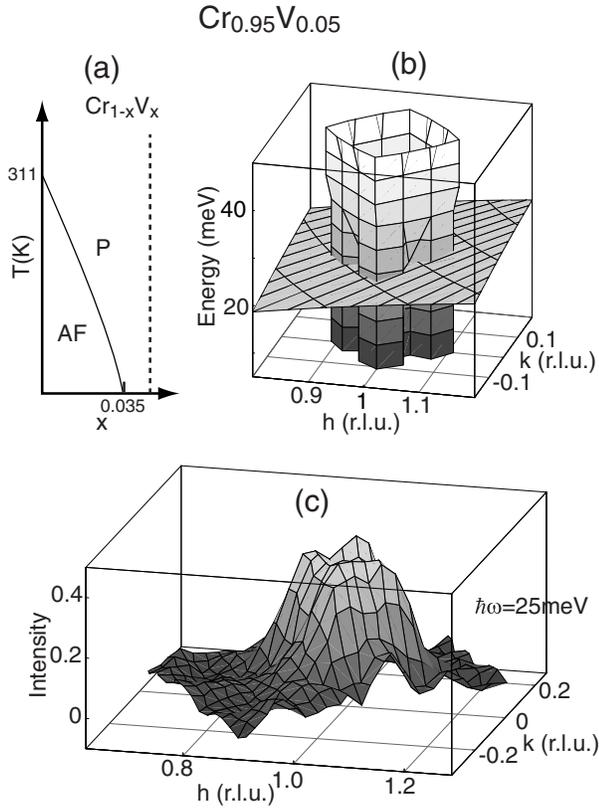}}
\caption[]{ 
(a) The magnetic phase diagram of the Cr$_{1-x}$V$_{x}$ system showing
the incommensurate antiferromagnet (AF) and paramagnetic (P) phases.
(b) Schematic representation of the dynamic susceptibility
$\chi^{\prime\prime}({\bf Q},\omega)$ (Eq.  \protect\ref{chi_expand})
of Cr$_{0.95}$V$_{0.05}$ with experimentally determined parameters.
For each energy, a contour at $1/2$ the maximum value of
$\chi^{\prime\prime}({\bf Q},\omega)$ is drawn, the shading represents
the maximum value.  For $E_i=35$ meV, $\chi^{\prime\prime}({\bf
Q},\omega)$ is sampled on the grey surface.  (c) Magnetic fluctuations
near (100) in Cr$_{0.95}$V$_{0.05}$ ($T$=12 K) measured using the HET
spectrometer. }
\label{HET25meV}
\end{figure}
\begin{figure}[t]
\epsfxsize=8cm
\epsffile{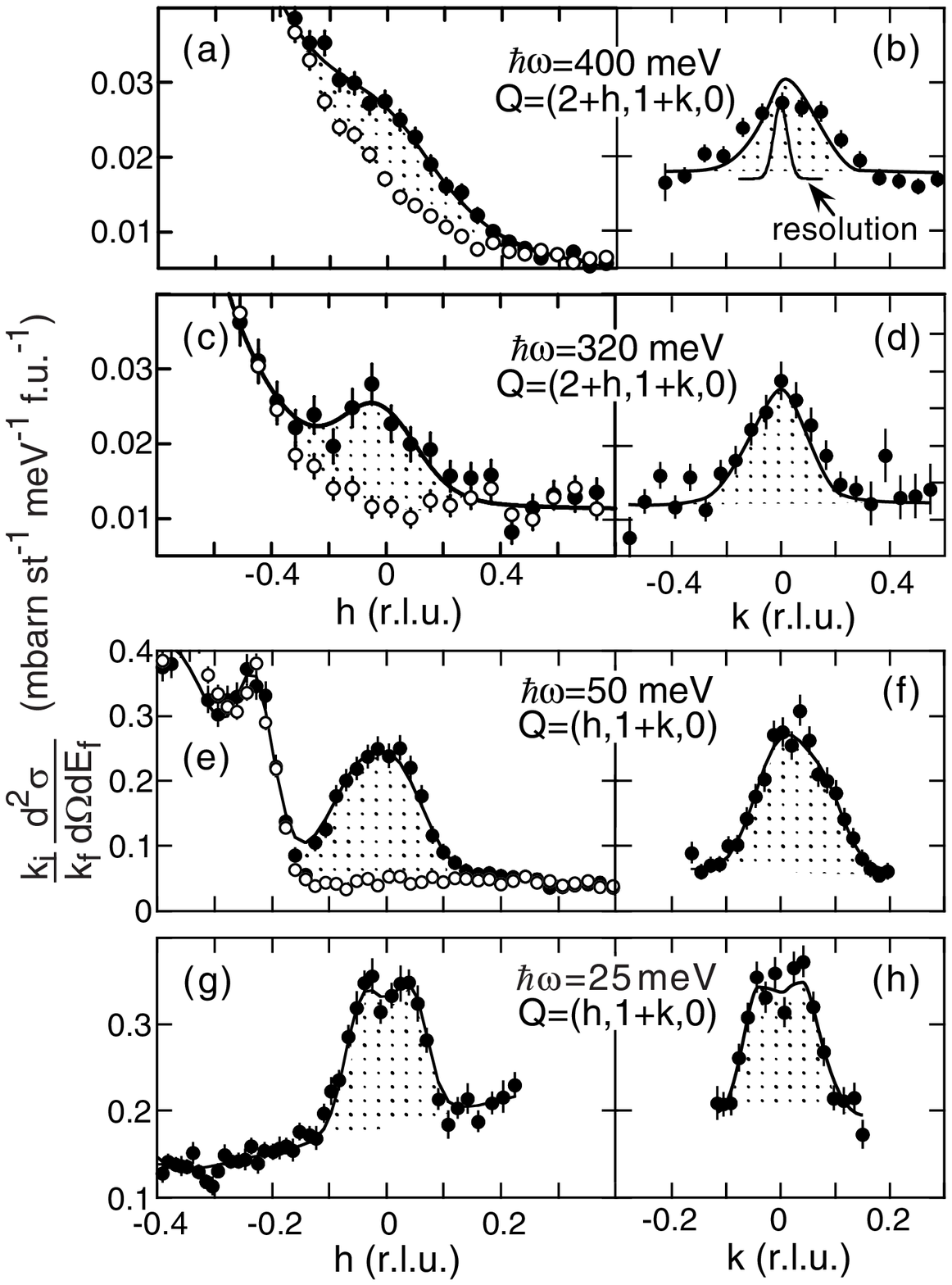}
\caption{
Magnetic scattering near the (100) and (210) reciprocal space
positions.  By increasing the incident neutron energy, we are able to
cut through the `column' of scattering depicted in
Fig. \protect\ref{HET25meV}(b) at different energy transfers.
The left and right panels show cuts in two perpendicular reciprocal
space directions.  Open circles are a
background measured at the same $\omega$ and $|{\bf Q}|$ but away from
the H position and the shaded area is the magnetic signal.
The incident energies were
(from bottom) E$_{i}$=35, 110, 705, 860 meV. Solid lines are
resolution-corrected fits of Eq. \protect\ref{chi_fit} to the data.}
\label{cuts}
\end{figure}

We first characterized the inelastic scattering at lower
$\omega$.  Previous measurements\cite{Fawcett88} have established that
for $\hbar\omega \approx 10$ meV, the magnetic response is peaked at
the six equivalent positions ${\bf Q}_{\delta}=(1\pm\delta,0,0)$,
$(1,\pm\delta,0)$ and $(1,0,\pm\delta)$.  For an incident neutron
energy $E_{i}$=35 meV, we probe the excitations in the $(h,k,0)$ plane
near (100) for $\hbar \omega \approx$ 25 meV as shown in
Fig.~\ref{HET25meV}(b).  Data collected under these conditions are
shown in Fig.~\ref{HET25meV}(c).  The response is a
diffuse structure in ${\bf Q}$ peaked near $(1 \pm \delta,0,0)$ and
$(1,\pm\delta,0)$.  One might visualize the overall magnetic response
in Cr$_{0.95}$V$_{0.05}$ as a `column' of scattering in ${\bf
Q}-\hbar\omega$ space as illustrated schematically in Fig.
\ref{HET25meV}(b).  By varying the incident energy $E_i$, we cut the
`column' of scattering at different $\omega$.  Fig.~\ref{cuts} shows
cuts along two perpendicular wavevector directions.  At all $\omega$
investigated, we observe a diffuse feature near the H position.  The
highest $\omega$ investigated was $\hbar\omega$=400 meV, although, the
excitations probably exist up to even higher $\omega$.  As the energy
transfer increases the scattering broadens in wavevector and only one
broad peak is observed at the highest $\omega$ investigated.  For all
energies the response is significantly broader than the
wavevector resolution.
%

A number of response functions have been used to
describe nearly antiferromagnetic metals 
\cite{Moriya70,Sato74,Millis90,Noakes90,Zha96,Aeppli97}. 
Following Zha et al. \cite{Zha96} for 
La$_{2-x}$Sr$_x$CuO$_4$, we use the function
\begin{eqnarray}
\label{chi_expand}
\chi({\bf Q},\omega) & = & \sum_{{\bf Q}_{\delta}}
\chi_{\delta} \left[ 1 + \frac{|{\bf Q-Q}_{\delta}|^2}{\kappa^{2}_{0}} 
- i \frac{\omega}{\omega_{\rm SF}} \right]^{-1},
\end{eqnarray}
where the sum is over the six ${\bf Q}_{\delta}$  positions surrounding the 
H point of the BZ.  Eq.~\ref{chi_expand} may be justified by expanding the 
Lindhard functions near the Fermi energy\cite{Moriya70} and thus 
should describe a Fermi liquid.  It also has a more general phenological 
justification in that it represents the
first symmetry-allowed terms in a power series expansion of 
$\chi^{-1}({\bf Q},\omega)$.
The observed scattering is related to the imaginary part of
Eq.~\ref{chi_expand} which is,
\begin{eqnarray}
\label{chi_im}
\chi^{\prime\prime}({\bf Q},\omega) & = &
\sum_{{\bf Q}_{\delta}}
\frac{\chi_{\delta} \kappa_0^4 \left[ \omega/\omega_{\rm SF} \right]}
{\left[ \kappa_0^2 + ({\bf Q}-{\bf Q}_{\delta})^2 \right]^2
+ \left[ \omega/\omega_{\rm SF} \right]^2 \kappa_0^4}.
\end{eqnarray}
The physical meaning of the parameters in Eq.~\ref{chi_im} is now clear. 
$\chi_{\delta}$ determines the overall amplitude of the response. 
In the low-frequency limit $\omega \ll \omega_{\rm SF}$,
the position and sharpness of the peaks in wavevector are determined by
$\delta$ and $\kappa_0$ respectively.  We find these parameters to
be $\delta$=0.08$\pm$0.01 r.l.u. and $\kappa_0$=0.11$\pm$0.01\AA$^{-1}$ from
fitting low-frequency ($\hbar\omega$=4~meV) data 
collected on a cold-source 
triple axis spectrometer.  Finally, the frequency dependence of the response 
i.e. how the peak widths and amplitudes change with frequency 
is controlled by the term 
$\left( \omega/\omega_{\rm SF}\right)^2\kappa^{4}_{0}$, where 
$\omega_{\rm SF}$ is the characteristic frequency over which changes occur.

To test whether Eq.~\ref{chi_im} describes the evolution of our data
with $\omega$, we fitted cuts at each $\omega$ using the general form 
\begin{eqnarray}
\label{chi_fit}
\chi^{\prime\prime}({\bf Q},\omega) & = &
\sum_{{\bf Q}_{\delta}}
\frac{\chi_{\delta}^{\prime\prime}(\omega)
[\kappa^4_0+\kappa^4_1(\omega)]}
{\left[\kappa^2_0+({\bf Q}-{\bf Q}_{\delta})^2\right]^2 +
\kappa^4_1(\omega)},
\end{eqnarray}
where the fitted $\omega$-dependent parameters 
$\chi^{\prime\prime}_{\delta}(\omega)$ and $\kappa_1(\omega)$ 
control the the height and sharpness of the incommensurate peaks 
respectively. The parameters $\delta$ and $\kappa_0$ were fixed 
at the values given above.  If Eq.~\ref{chi_im}
describes the response, comparing Eq.~\ref{chi_im} and
Eq.~\ref{chi_fit} predicts that the fitted parameters will vary as,
\begin{eqnarray}
\label{chi_p_eqn}
\chi^{\prime\prime}_{\delta}(\omega)&=&
\frac{\chi_{\delta} \left[ \omega / \omega_{\rm SF} \right] }
{1+ \left[ \omega^2 /\omega_{\rm SF}^2 \right] }
\end{eqnarray}
and 
\begin{eqnarray}
\label{kappa}
\kappa_1^{2} (\omega) & = & \frac{\omega}{\omega_{\rm SF}} \kappa_0^2.
\end{eqnarray}   

\begin{figure}[t]
\epsfxsize=7cm
\epsfclipon
\epsffile{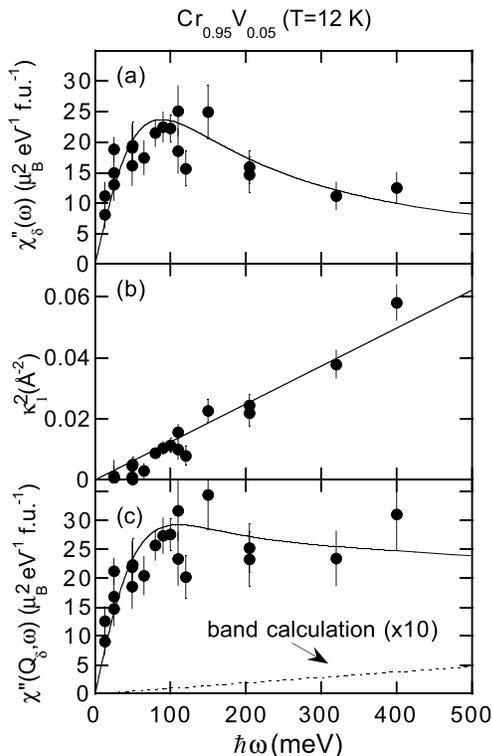}
\caption{ 
The energy dependence of: (a) the amplitude parameter
$\chi^{\prime\prime}_{\delta}(\omega)$; (b) the sharpness parameter 
$\kappa_1^2(\omega)$ ; (c) the susceptibility at the ${\bf Q}_{\delta}$
position $\chi^{\prime\prime}({\bf Q}_{\delta},\omega)$.
}
\label{chi_p_fig}
\end{figure}

The solid lines in Fig.~\ref{cuts} correspond to resolution-corrected
fits of Eqs.~\ref{chi_fit}-\ref{kappa} for different $\omega$.
Fig.~\ref{chi_p_fig}(a) and (b) show the resulting 
$\chi^{\prime\prime}_{\delta}(\omega)$ and 
$\kappa_1^2(\omega)$ values for all data collected.  
The solid line in Fig.~\ref{chi_p_fig}(a)
is a fit of Eq.~\ref{chi_p_eqn} which yields values for the spin
fluctuation energy and amplitude parameters of
$\hbar\omega_{\rm SF}$=88$\pm$10 meV and 
$\chi_{\delta}$=45$\pm$3~$\mu_{B}^{2} {\rm eV}^{-1} {\rm f.u.}^{-1}$
respectively. Fig.~\ref{chi_p_fig}(b) shows that
the width parameter $\kappa_1^2$ displays an approximately
linear variation with $\omega$ as suggested by Eq.~\ref{kappa}. 
The gradient of the fitted line 
$\kappa_0^2/\omega_{\rm SF}$=$0.118\pm0.006$ \AA$^{-2}$ eV$^{-1}$ 
gives a second estimate of $\hbar\omega_{\rm SF}$=102$\pm$24 meV.  
Obtaining two indistinguishable estimates of $\omega_{\rm SF}$ in this way
demonstrates that Eq.~\ref{chi_expand} provides a good description of 
our data:  the $\omega$ dependence of the
the amplitude and sharpness of the peaks are controlled by the parameter
$\omega_{\rm SF}$.  
%

The starting point for models of the dynamic response of chromium and
its alloys is the Fermi surface.  It is believed that 
the peak in the dynamic susceptibility is due to the imperfect nesting
of an electron `jack' and a larger hole `octahedron'.  Staunton {\it et
al.} \cite{Staunton99} have recently performed {\it ab-initio} calculations
of the spin susceptibility for Cr$_{0.95}$V$_{0.05}$. To compare our 
results with these and other theoretical models, we have evaluated the 
$\omega$ dependence of $\chi^{\prime\prime}({\bf Q}_{\delta},\omega)$ 
for each $\omega$ probed assuming Eq.~\ref{kappa}.
The resulting response is shown in Fig.~\ref{chi_p_fig}(c).
$\chi^{\prime\prime}({\bf Q}_{\delta},\omega)$ differs from
$\chi^{\prime\prime}_{\delta}(\omega)$  because
Eq.\ (3) sums contributions centered at six equivalent and nearby
wavevectors ${\bf Q}_{\delta}$.
The dashed line in Fig.~\ref{chi_p_fig}(c) shows the non-interacting 
susceptibility $\chi_{0}^{\prime\prime}({\bf Q},\omega)$
calculated from the band structure and neglecting the exchange
interaction i.e.\ the bare Lindhard function \cite{Staunton99}. 
Within a simple RPA model, the interacting
susceptibility $\chi({\bf Q}_{\delta},\omega)$ is given by 
$\chi^{-1}({\bf Q},\omega)=\chi_{0}^{-1}({\bf Q},\omega)-\lambda$,
where $\lambda$ is the mean field parameter describing the Coulomb
interaction.
In the low $\omega$ limit 
$\chi^{\prime\prime}({\bf Q},\omega) \propto \omega$ 
and,
\begin{eqnarray}
\label{chi_low_w}
\chi^{\prime\prime}({\bf Q},\omega) & = & 
\frac{\chi^{\prime\prime}_{0}({\bf Q},\omega)}
{\left[1-\lambda \chi^{\prime}_{0}({\bf Q},\omega)\right]^2}.
\end{eqnarray}
Comparing the initial gradients
of the two lines in Fig.~\ref{chi_p_fig}(c), we estimate the exchange 
enhancement factor 
$\left[1-\lambda \chi^{\prime}_{0}({\bf Q}_{\delta},0) \right]^{-1}$ =
28$\pm$4, demonstrating
that Cr$_{0.95}$V$_{0.05}$ is indeed very strongly exchange enhanced.  

%
\begin{figure}[t]
\epsfxsize=7cm
\epsffile{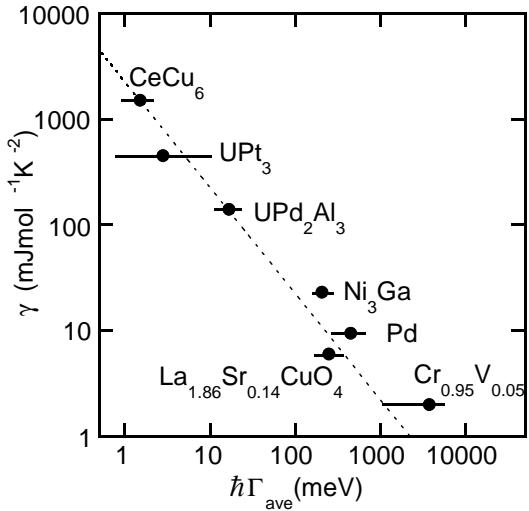}
\caption{ The coefficient $\gamma$ of the low temperature electronic 
specific heat $\gamma T$ is plotted again the wavevector-averaged spin 
relaxation rate $\Gamma_{\text{ave}}$, where 
$\Gamma^{-1}_{\text{ave}}=\left< \Gamma^{-1}({\bf Q})
\right>_{\text{BZ}}$.
$\Gamma_{\text{ave}}$ has been computed from data in 
Refs.~\protect\cite{Gamma_refs}, horizontal bars are estimates of the 
uncertainty in $\Gamma_{\text{ave}}$. The dotted line is 
Eq.~\protect\ref{gamma}.}
\label{gammaGamma}
\end{figure}
Many systems displaying strong magnetic fluctuations show dramatically
renormalized thermodynamic properties. For example, heavy fermions
exhibit a very large electronic linear heat capacity at low
temperature. In spite of its large exchange enhancement,
Cr$_{0.95}$V$_{0.05}$ has a relatively small electronic specific heat
$\gamma T$, $\gamma$=2 mJK$^{-2}$ \cite{Heiniger66}. To understand why
the quasiparticles in Cr$_{0.95}$V$_{0.05}$ are not heavier, we must
estimate the magnetic contribution to $\gamma$.  If the
excitations can be described as a set of overdamped oscillators with
wavevector-dependent relaxation rate (characteristic
frequency) $\Gamma({\bf Q})$, then 
\cite{Edwards92,Moriya95,note1},
\begin{eqnarray}
\label{gamma}
\gamma &=&\frac{\pi k_{B}^{2}}{\hbar} \frac{1}{\Gamma_{\rm ave}}=
\frac{\pi k_{B}^{2}}{\hbar} \left< \frac{1}{\Gamma({\bf Q})} 
\right>_{\rm BZ},
\end{eqnarray}
and $\left< \right>$ denotes a Brillouin-zone average. To put our results into
a more general context, we have evaluated $\Gamma_{\text{ave}}$ from
Eq.~\ref{gamma} for Cr$_{0.95}$V$_{0.05}$ and a number of other
materials where the wavevector dependence of the magnetic response is
known throughout a large part of reciprocal space.
Fig.~\ref{gammaGamma} shows the results plotted against $\gamma$.  The
universal relationship Eq.~\ref{gamma} holds across widely differing
systems such as $3d$ transition metals, their oxides, and heavy
fermion compounds, demonstrating that the spin channel dominates the
electronic entropy in all of these systems.  

One might view the materials in Fig.~\ref{gammaGamma} as being close
to a QCP.  Each of the materials has `soft' magnetic excitations
in some region of reciprocal space.  In spite of the large enhancement
in Cr$_{0.95}$V$_{0.05}$, the magnetic fluctuations in a relatively 
small portion of reciprocal space (where $\Gamma$ is small) contribute 
to $\gamma$. In contrast, the heavy fermion systems have large $\gamma$ 
because the spin fluctuations are soft (have low $\omega$)
over larger portions of reciprocal space. For predominately antiferromagnetic 
fluctuations, $\gamma$ scales
roughly as $\kappa_{0}^{d}/\omega_{SF}$, where $d$ is the system 
dimensionality. Thus, to raise $\gamma$, it is advantageous to lower $d$ and
$\omega_{\text{SF}}$ and increase $\kappa_0$. Heavy fermion systems have 
larger $\kappa_0$ and smaller $\omega_{\text{SF}}$ than Cr$_{0.95}$V$_{0.05}$ 
and hence larger $\gamma$ values.

In summary, we have measured the high-frequency dynamic magnetic
response of the paramagnetic alloy Cr$_{0.95}$V$_{0.05}$ which is
close to incommensurate magnetic order and has a large exchange 
enhancement.  We observe
strong magnetic correlations at epithermal energies up to 400 meV.  
The observed response can be described by a
remarkably-simple modified-MMP function where the $\omega$ dependence of
the response is controlled by a single parameter.  We have computed the
BZ average of the magnetic relaxation rate for 
Cr$_{0.95}$V$_{0.05}$ and other materials close to magnetic
order (or a QCP), this demonstrates the relationship
between this quantity and the low-temperature quasiparticle specific
heat, and also account for why Cr$_{0.95}$V$_{0.05}$, in spite of its
large exchange enhancement, is not a heavy fermion system.

We grateful for helpful discussions with P. W. Michell, J. Lowden, 
R. Fishman, J. Staunton and O. Stockert.


\end{document}